\documentclass[prl,aps,twocolumn,showpacs]{revtex4}
\usepackage{epsfig}
\newcommand{\ket}[1]{|#1\rangle}
\newcommand{\bra}[1]{\langle #1|}
\newcommand{\Tr}{\text{Tr}}
\begin{document}
\title{Off-diagonal geometric phase for mixed states}
\author{Stefan Filipp$^{1,2}$\footnote{Electronic address: 
stefan@copilot.at} and Erik Sj\"{o}qvist$^{1}$\footnote{
Electronic address: erik.sjoqvist@kvac.uu.se}} 
\affiliation{$^{1}$Department of Quantum Chemistry, Uppsala University, 
Box 518, Se-751 20 Uppsala, Sweden \\ 
$^{2}$Institut f\"{u}r Theoretische Physik, Vienna 
University of Technology, Wiedner Hauptstrasse 8-10/136, A-1040 
Vienna, Austria.}
\date{\today}
\begin{abstract}
We extend the off-diagonal geometric phase [Phys. Rev. Lett. 
{\bf 85}, 3067 (2000)] to mixed quantal states. The 
nodal structure of this phase in the qubit (two-level) case 
is compared with that of the diagonal mixed state geometric 
phase [Phys. Rev. Lett. {\bf 85}, 2845 (2000)]. Extension 
to higher dimensional Hilbert spaces is delineated. A physical 
scenario for the off-diagonal mixed state geometric phase in 
polarization-entangled two-photon interferometry is proposed.
\end{abstract}
\maketitle
The geometric phase discovered by Berry \cite{berry84} for 
cyclic adiabatic evolution has led to important insights 
into the geometry of quantal evolution as well as to   
several generalizations. Extension to the nonadiabatic 
cyclic case was given by Aharonov and Anandan \cite{aharonov87}, 
who pointed out that the geometric phase is due to the curvature 
of the quantal state space. Based upon Pancharatnam's \cite{pancharatnam56} 
work on interference of light in distinct state of polarization, 
Samuel and Bhandari \cite{samuel88} provided a general setting 
for the geometric phase so as to cover noncyclic and 
nonunitary evolutions. These noncyclic concepts become undefined 
when the interfering states are orthogonal, which led Manini and 
Pistolesi \cite{manini00} to introduce the off-diagonal geometric 
phase for pure states in adiabatic evolution. This adiabaticity 
assumption was subsequently removed by Mukunda {\it et al.} 
\cite{mukunda02} and the off-diagonal pure state geometric 
phase was verified by Hasegawa {\it et al.} \cite{hasegawa01} 
in a neutron experiment. 

Another development of the geometric phase has been its 
extension to the mixed state case. Uhlmann \cite{uhlmann86} 
was probably first to address this issue in the context of 
purification. More recently another mixed state geometric phase 
was discovered in the experimental context of interferometry 
by Sj\"{o}qvist {\it et al.} \cite{sjoqvist00}. It has been 
pointed out \cite{bhandari01} that this latter mixed 
state geometric phase can be undefined at nodal points  
in the parameter space where the interference visibility 
vanishes. 

In this Letter, we expand the concept of mixed state geometric 
phase to the off-diagonal case. This off-diagonal mixed state 
geometric phase could contain interference information 
when the ``diagonal'' phase in \cite{sjoqvist00} is undefined. 
The off-diagonal mixed state geometric phase reduces 
to that proposed in \cite{manini00} in the limit of pure states 
and it may be verified experimentally as a shift in the interference 
oscillations in a polarization-entangled two-photon interferometry 
set up.  

The idea behind the off-diagonal pure state geometric phase 
arises when considering parallel transport generated by the 
operator $U^{\parallel}$ of the $j^{th}$ eigenstate $\ket{\psi_j}$ 
of some time-independent Hermitian operator along the path 
$\Gamma$ in state space to the $k^{th}$ eigenstate 
$\ket{\psi_k} = U^{\parallel} \ket{\psi_j}$. Then the scalar 
product $\bra{\psi_j} U^{\parallel} \ket{\psi_j}$ vanishes 
and the concomitant relative phase becomes undefined. The 
only phase information left is in the cross scalar product 
$\bra{\psi_k} U^{\parallel} \ket{\psi_{j}} \ (j\neq k)$, 
from which the off-diagonal geometric phase factor 
$\gamma_{jk}^{\Gamma}$ can defined as 
\begin{equation}
\label{eq:man5}
\gamma_{jk}^{\Gamma} \equiv 
\sigma_{jk} \sigma_{kj}
\end{equation}
where $\sigma_{jk} = \Phi [\bra{\psi_j} U^{\parallel} 
\ket{\psi_k}]$ with $\Phi [z] = z/|z|$. This quantity  
is gauge invariant and consequently measurable. Furthermore, 
it is reparametrization invariant, real-valued, and it is 
solely a property of the subjacent geometry of state 
space. In the qubit (two-level) case it can be shown that 
the off-diagonal geometric phase $\arg \gamma_{jk}^\Gamma$ 
becomes $\pi$ for any open path $\Gamma$ on the Bloch sphere.  

The above can be generalized to $l$ mutually orthogonal states
by defining  
\begin{eqnarray}
\label{eq:man6}
\gamma_{j_1j_2 \ldots j_l}^{(l)\Gamma} \equiv  
\sigma_{j_{1}j_{2}} \sigma_{j_{2}j_{3}} \ldots 
\sigma_{j_{l}j_{1}}  
\end{eqnarray}
as any cyclic product of $\sigma$'s is gauge invariant.  
If $l=1$ this reduces to the diagonal geometric phase factor
$\gamma_j^\Gamma$ and if $l=2$ we obtain the off-diagonal pure 
state geometric phase. For $l>2$ more complex phase relations 
among off-diagonal components of the eigenstates at the end-points 
of $\Gamma$ can be described. Such phase relations have been 
analyzed \cite{manini00,pistolesi00} for the deformed microwave 
resonator experiments in \cite{lauber94}. 

Due to decoherence effects or improper preparation procedures 
it is more realistic to talk about mixed states in quantum 
mechanics. To cover such situations the concept of diagonal 
mixed state geometric phase was introduced in \cite{sjoqvist00} 
by considering Mach-Zehnder interferometry with a nondegenerate 
mixed internal input state $\rho$. This phase arises naturally 
as the shift of the interference oscillations determined by 
\begin{equation} 
\gamma_{\rho} = 
\Phi \big[ \Tr \big( U^{\parallel} \rho \big) \big] 
\label{eq:mixeddiag}
\end{equation}
with the unitarity $U^{\parallel}$ parallel transporting each 
eigenstate of $\rho$ in one arm of the interferometer. 
The geometric phase factor $\gamma_{\rho}$ is a property of 
the subjacent geometry of state space and reduces to that of 
the standard geometric phase in the limit of pure states. It 
becomes undefined at its nodal points where the  visibility 
factor $\big|\Tr \big( U^{\parallel} \rho \big)\big|$ 
vanishes \cite{bhandari01}.  

To generalize the above to the off-diagonal mixed state case 
we have to find an appropriate notion of ``maximal orthogonality'' 
between unitarily connected density matrices. One approach 
would be to take the density matrices $\rho$ and $\rho' = 
U \rho U^{\dagger}$ as maximally orthogonal if their Bures 
fidelity \cite{bures69} ${\cal F}_B [\rho ,\rho' ] = \big[ 
\Tr \sqrt{\sqrt{\rho}\rho'\sqrt{\rho}} \big]^2$ is at infimum, 
as this would ensure maximal distinguishability between 
$\rho$ and $\rho'$ \cite{jozsa94}. However, for our needs 
a simpler definition is adequate, namely to say that $\rho$ 
and $\rho'$ are ``quasi-orthogonal'' if 
\begin{eqnarray} 
\rho & = & \sum_{k=1}^{N} \lambda_{k} \ket{\psi_{k}} 
\bra{\psi_{k}} , 
\nonumber \\ 
\rho' & \equiv & \rho^{\perp} =  
\sum_{k=1}^{N} \lambda_{k} \ket{\psi_{k}^{\perp}} 
\bra{\psi_{k}^{\perp}}
\label{eq:orthogonality}  
\end{eqnarray} 
with $\bra{\psi_{k}} \psi_{k}^{\perp} \rangle = 0$ for 
each $k = 1,\ldots,N$, $N$ being the dimension of the Hilbert 
space. In the qubit case, this notion is equivalent to 
solving the minimization problem for the Bures fidelity, but 
in higher dimensional cases it is straightforward to find 
examples where the two approaches differ. 

Using the above concept of quasi-orthogonality, we now 
define the off-diagonal mixed state geometric phase 
factor $\gamma_{\rho\rho^\perp}$ for nondegenerate 
density matrices as 
\begin{equation}
\label{eq:mixdef1}
\gamma_{\rho\rho^\perp} \equiv 
\Phi \big[ \Tr \big( U^{\parallel} \sqrt{\rho} U^{\parallel} 
\sqrt{\rho^\perp} \big) \big] , 
\end{equation}
where the unitarity $U^{\parallel}$ parallel transports 
each eigenstate $\ket{\psi_k}$ of $\rho$. This definition 
can be seen as a natural extension of \cite{manini00} as it 
reduces to Eq. (\ref{eq:man5}) in the limit of pure states. 
Furthermore, it may be well-defined for points in parameter 
space, where the mixed state diagonal geometric phase factor is
undefined and it is manifestly gauge invariant under a phase 
transformation of the spectral bases $\{ \ket{\psi_k} \}$ 
and $\{ \ket{\psi_k^{\perp}} \}$ of $\rho$ and $\rho^{\perp}$. 
We demonstrate below that it can be assigned an operational 
meaning in terms of a purification lift that can be 
experimentally tested using two-particle interferometry. 

We may extend the off-diagonal mixed state geometric phase  
to $l\leq N$ mutually quasi-orthogonal density matrices 
$\rho_{j_k}$, $k=1,2,\ldots ,l$, yielding the expression 
\begin{equation}
\label{eq:genoffdiag}
\gamma_{\rho_{j_1}\rho_{j_2}\ldots\rho_{j_l}}^{(l)} 
\equiv \Phi \big[ \Tr \big( U^{\parallel} \sqrt[l]{\rho_{j_1}} 
U^{\parallel} \sqrt[l]{\rho_{j_2}} \ldots U^{\parallel} 
\sqrt[l]{\rho_{j_l}} \big) \big] ,  
\end{equation}
which is gauge invariant and independent of cyclic permutations 
of the indexes $j_{1},j_{2}, \ldots j_{l}$. It reduces to the 
diagonal mixed state geometric phase $\arg\Tr[U^{\parallel} \rho]$ 
for $l=1$ and to the off-diagonal mixed state geometric phase 
$\arg\Tr [U^{\parallel} \sqrt{\rho} U^{\parallel} \sqrt{\rho^\perp}]$ 
for $l=2$. In the limit of pure states  it is equivalent to 
Eq. (\ref{eq:man6}). 

To delineate the nodal structure of the mixed state geometric 
phases in Eqs. (\ref{eq:mixeddiag}) and (\ref{eq:mixdef1}) let 
us first consider the qubit case with $\rho = \frac{1}{2} 
(1+r\sigma_{z})$, where $r\neq 0$ is the length of the Bloch vector 
and $\sigma_{z}$ is the standard Pauli operator in the 
$\ket{\psi_1},\ket{\psi_2}$ basis. Using the definition of 
quasi-orthogonality yields $\rho^{\perp} = \frac{1}{2} 
(1-r\sigma_{z})$. Putting these density matrices into Eq. 
(\ref{eq:mixdef1}) we obtain  
\begin{eqnarray}
\Tr \big( U^{\parallel} \sqrt{\rho} U^{\parallel} 
\sqrt{\rho^\perp} \big) & = &  
\eta^{2} \sqrt{{\cal F}_{B} [\rho ,\rho^{\perp}]} \cos \Omega 
\nonumber \\ 
 & & + (1-\eta^{2}) \gamma_{12}^{\Gamma} .
\label{eq:offqubit}
\end{eqnarray}
Here, $\eta = \big|\bra{\psi_1} U^{\parallel} \ket{\psi_1}\big|$ 
is the pure state visibility and $\Omega$ is the solid angle 
enclosed by the path $\Gamma$ and the shortest geodesic connecting 
its end-points on the Bloch sphere. Similarly, the expression 
in the diagonal case becomes \cite{sjoqvist00} 
\begin{eqnarray}
\Tr \big( U^{\parallel} \rho \big) & = & \eta 
\sqrt{\cos^{2} \frac{\Omega}{2} + r^{2} \sin^{2} \frac{\Omega}{2}} 
\nonumber \\ 
 & & \times \exp \Big( - i\arctan \big[ r\tan \frac{\Omega}{2} 
\big] \Big) . 
\label{eq:qubit}
\end{eqnarray} 

\begin{figure}[ht!]
\begin{center}
\includegraphics[width=8 cm]{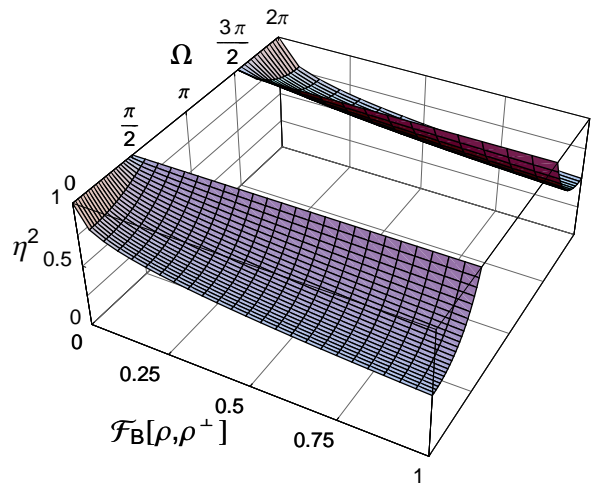}
\end{center}
\caption{Nodal surfaces of the off-diagonal mixed state geometric 
phase for a qubit. For Bures fidelity ${\cal F}_{B}>0$ (mixed states), 
there are nodes also for paths with pure state visibility 
$\eta \neq 1$ at various solid angles $\Omega$.}
\end{figure}

The Bures fidelity ${\cal F}_{B} [\rho , \rho^{\perp}] = 
1-r^{2}$ vanishes for pure states and the off-diagonal 
geometric phase becomes undefined only for cyclic evolution, 
where the diagonal geometric phase is well-defined. 
In the mixed state case $0<r<1$ the diagonal geometric phase 
becomes undefined only for rotations that flip the Bloch vector, 
corresponding to $\eta = 0$. For such rotations 
$\big| \Tr \big( U^{\parallel} \sqrt{\rho} 
U^{\parallel} \sqrt{\rho^\perp} \big)\big| = 1$ and the 
off-diagonal mixed state geometric phase is well-defined 
and equals the pure state value $\arg \gamma_{12}^{\Gamma} = 
\pi$. This shows in the qubit case that the diagonal and 
off-diagonal mixed state geometric phases never become 
undefined simultaneously. 

The off-diagonal mixed state geometric phase in the qubit 
case has a nontrivial nodal structure that arises due to 
the nonvanishing Bures fidelity. This can be seen by putting 
the left-hand side of Eq. (\ref{eq:offqubit}) to zero and 
solving for $\eta^2$ yielding 
\begin{equation} 
\eta^{2} = \big( 1+\sqrt{{\cal F}_{B} [\rho,\rho^{\perp}]} 
\cos \Omega \big)^{-1} , 
\label{eq:qubitnodal}
\end{equation}
which has solutions at $\eta \neq 1$ for $\cos \Omega >0$ 
and ${\cal F}_{B} [\rho,\rho^{\perp}] >0$. Thus, the 
off-diagonal mixed state geometric phase factor may 
change sign across the nodal surfaces in the parameter space 
$({\cal F}_{B} [\rho,\rho^{\perp}],\eta,\Omega)$ defined by 
the solutions of Eq. (\ref{eq:qubitnodal}), as shown in 
Fig. 1. Thus, the corresponding off-diagonal mixed state 
geometric phase can take both values $0$ and $\pi$, contrary 
to the corresponding pure state phase, which can only be $\pi$. 

In the maximally mixed state case $(r=0)$, the density 
matrix is degenerate and the geometric phases are undefined 
since there is no direction in space singled out. Still, 
there is a unique notion of relative phase in this case 
with a nontrivial nodal structure discussed in \cite{bhandari01}. 
The generic situation is covered by the unitarity 
$U=e^{-i\delta \sigma_{z}}$ from which we obtain 
$\Tr [U\sqrt{\rho} U \sqrt{\rho^{\perp}}] = \cos 2\delta$ 
and $\Tr [U \rho ] = \cos \delta$ with no common nodal points.   

Next, let us generalize to arbitrary Hilbert space dimensions 
$N$. We take the set 
\begin{eqnarray} 
\rho_{1} & = & \lambda_{1} \ket{\psi_1} \bra{\psi_1} + 
\lambda_{2} \ket{\psi_2} \bra{\psi_2} + \ldots + 
\lambda_{N} \ket{\psi_N} \bra{\psi_N} , 
\nonumber \\ 
\rho_{2} & = & \lambda_{1} \ket{\psi_2} \bra{\psi_2} + 
\lambda_{2} \ket{\psi_3} \bra{\psi_3} + \ldots +  
\lambda_{N} \ket{\psi_1} \bra{\psi_1} , 
\nonumber \\ 
 & \ldots & 
\nonumber \\ 
\rho_{N} & = & \lambda_{1} \ket{\psi_N} \bra{\psi_N} + 
\lambda_{2} \ket{\psi_1} \bra{\psi_1} + \ldots 
\nonumber \\ 
 & & + \lambda_{N} \ket{\psi_{N-1}} \bra{\psi_{N-1}} . 
\label{eq:Nstates}
\end{eqnarray}
of mutually quasi-orthogonal nondegenerate density matrices 
and consider parallel transporting unitarities that permute 
the eigenstates $\ket{\psi_1}, \ldots , \ket{\psi_N}$. 
By appropriate labeling of the eigenstates, such operators 
can always be decomposed into the direct sum
\begin{equation}
U^{\parallel} = u_p^{\parallel} \oplus u_d^{\parallel} 
\end{equation}
where $u_p^{\parallel}$ permutes $\ket{\psi_1} \rightarrow 
\ket{\psi_m} \rightarrow \ldots \rightarrow \ket{\psi_2}
\rightarrow \ket{\psi_1}$ and $u_d^{\parallel}$ is diagonal 
in the remaining $N-m$ eigenstates. These terms do not mix 
so that one may write   
\begin{eqnarray} 
\Tr \big( U^{\parallel} \sqrt[l]{\rho_{j_1}} \ldots 
U^{\parallel} \sqrt[l]{\rho_{j_l}} \big) =  
{\cal P}_{\rho_{j_1}\ldots\rho_{j_l}}^{(l)} + 
{\cal D}_{\rho_{j_1}\ldots\rho_{j_l}}^{(l)} , 
\end{eqnarray}
where ${\cal P}_{\rho_{j_1}\ldots\rho_{j_l}}^{(l)} = 
\Tr \big( u_p^{\parallel} \sqrt[l]{\rho_{j_1}} \ldots 
u_p^{\parallel} \sqrt[l]{\rho_{j_l}} \big)$ and   
${\cal D}_{\rho_{j_1}\ldots\rho_{j_l}}^{(l)} =  
\Tr \big( u_d^{\parallel} \sqrt[l]{\rho_{j_1}} 
\ldots u_d^{\parallel} \sqrt[l]{\rho_{j_l}} \big)$. If 
$l=K\times m$, $K$ integer $\leq N/m$ and $m\geq 2$, then 
\begin{eqnarray}  
{\cal P}_{\rho_{j_1}\ldots\rho_{j_l}}^{(l)} = 
\big[ (-1)^{m-1} \det u_p^{\parallel} \big]^K 
f_{\rho_{j_1}\ldots\rho_{j_l}}^{(l)} 
(\lambda_1,\ldots,\lambda_N) , 
\end{eqnarray}
where the $f^{(l)}$'s can be written as sums of terms of the 
form $\sqrt[l]{\lambda_{a_1} \ldots \lambda_{a_l}}$. For other 
$l$, the ${\cal P}^{(l)}$'s vanish as there is $K\times m$ steps 
needed to connect $\sqrt[l]{\rho_{j_1}}$ and 
$\sqrt[l]{\rho_{j_l}}$ with $u_p^{\parallel}$. In the extreme 
case where all $N$ 
eigenstates are permuted, only ${\cal P}_{\rho_{j_1} 
\ldots \rho_{j_l}}^{(N)}$ may be nonvanishing. In particular, 
we have the $\lambda$-independent expression 
${\cal P}_{\rho_{1} \rho_{2} \ldots \rho_{N}}^{(N)} = 
(-1)^{N-1} \det U^{\parallel}$, while the existence of each 
of the remaining $\gamma^{(N)}$'s depends upon the rank of 
the $\rho$'s. Turning to the contribution from the 
diagonal part of $U^{\parallel}$, we have 
\begin{eqnarray}
{\cal D}_{\rho_{j_1}\ldots\rho_{j_l}}^{(l)} = 
\sum_{k=m+1}^N \big( U_{kk}^{\parallel} \big)^l 
\sqrt[l]{\lambda_{k_1} \ldots \lambda_{k_l}} 
\label{eq:diagonal}  
\end{eqnarray}
with $U_{kk}^{\parallel}$ the matrix elements of 
$u_{d}^{\parallel}$ in the eigenbasis of the $\rho$'s. 
As the density matrices are nondegenerate, it follows that 
all $\lambda_{k_\alpha}$ are different in each term on the 
right-hand side of Eq. (\ref{eq:diagonal}) and ${\cal D}_{\rho_{j_1} 
\ldots\rho_{j_l}}^{(l)}$ must vanish if $l>$ rank of the 
$\rho$'s.  

Let us revisit the qubit $(N=2)$ case using the above general 
theory. If $m=0$, both $\gamma_{\rho_1}^{(1)}$ and 
$\gamma_{\rho_2}^{(1)}$ exist. Moreover, we have   
\begin{eqnarray} 
{\cal D}_{\rho_{1} \rho_{2}}^{(2)} = \sqrt{\lambda_1\lambda_2} 
\big[ \big( U_{11}^{\parallel} \big)^2 + 
\big( U_{22}^{\parallel} \big)^2 \big] , 
\end{eqnarray}
which is consistent with Eq. (\ref{eq:offqubit}) for $\eta=1$. 
In the permutation case $m=2$, we may use $(-1)^{N-1} 
\det U^{\parallel} = -1$ for $N=2$ and obtain 
\begin{eqnarray} 
{\cal P}_{\rho_{1} \rho_{2}}^{(2)} = -1,  
\end{eqnarray}
in agreement with Eq. (\ref{eq:offqubit}) for $\eta=0$. 

As a further illustration, let us work out the $N=3$ case 
in detail. For $m=0$, all the $\gamma^{(1)}$'s are well-defined. 
The dependence upon the rank of the density matrices is visible 
for higher $l$, namely      
\begin{eqnarray} 
{\cal D}_{\rho_{1} \rho_{2}}^{(2)} & = & \sqrt{\lambda_1 \lambda_3}  
\big( U_{11}^{\parallel} \big)^2 + \sqrt{\lambda_1 \lambda_2} 
\big( U_{22}^{\parallel} \big)^2  
\nonumber \\ 
 & & + \sqrt{\lambda_2 \lambda_3} 
\big( U_{33}^{\parallel} \big)^2 , 
\nonumber \\ 
{\cal D}_{\rho_{1} \rho_{2} \rho_{3}}^{(3)} & = & 
{\cal D}_{\rho_{1} \rho_{3} \rho_{2}}^{(3)} =  
\sqrt[3]{\lambda_1 \lambda_2 \lambda_3} 
\nonumber \\ 
 & & \times \big[ \big( U_{11}^{\parallel} \big)^3 + 
\big( U_{22}^{\parallel} \big)^3 + 
\big( U_{33}^{\parallel} \big)^3 \big] 
\end{eqnarray} 
with ${\cal D}_{\rho_{2} \rho_{3}}^{(2)}$ and 
${\cal D}_{\rho_{3} \rho_{1}}^{(2)}$ obtained by permutations 
of the $\lambda$'s. In the $m=2$ case, $\ket{\psi_1} \rightarrow 
\ket{\psi_2} \rightarrow \ket{\psi_1}$ while $\ket{\psi_3}$  
undergoes cyclic evolution. Explicitly we have   
\begin{eqnarray} 
{\cal D}_{\rho_{1}}^{(2)} & = & \lambda_3  U_{33}^{\parallel} ,  
\nonumber \\ 
{\cal D}_{\rho_{1} \rho_{2}}^{(2)} & = & 
\sqrt{\lambda_2 \lambda_3}  \big( U_{33}^{\parallel} \big)^2 , 
\nonumber \\ 
{\cal D}_{\rho_{1} \rho_{2} \rho_{3}}^{(3)} & = & 
{\cal D}_{\rho_{1} \rho_{3} \rho_{2}}^{(3)} =  
\sqrt[3]{\lambda_1 \lambda_2 \lambda_3} 
\big( U_{33}^{\parallel} \big)^3 , 
\nonumber \\ 
{\cal P}_{\rho_{1} \rho_{2}}^{(2)}
 & = & U_{12}^{\parallel} U_{21}^{\parallel} 
\big( \lambda_{1} + \sqrt{\lambda_2 \lambda_3} \big) , 
\end{eqnarray}
where we have used $(-1)^{m-1} \det u_{p}^{\parallel} = 
U_{12}^{\parallel} U_{21}^{\parallel}$. The remaining 
${\cal D}_{\rho_{2}}^{(1)}$, ${\cal D}_{\rho_{3}}^{(1)}$, 
${\cal D}_{\rho_{2} \rho_{3}}^{(2)}$, ${\cal D}_{\rho_{3} 
\rho_{1}}^{(2)}$, ${\cal P}_{\rho_{2} \rho_{3}}^{(2)}$, 
${\cal P}_{\rho_{3} \rho_{1}}^{(2)}$ are given by appropriate 
permutations of the $\lambda$'s. For $m=3$ (full 
permutation) the only possible contributions are 
\begin{eqnarray}
{\cal P}_{\rho_{1} \rho_{2} \rho_{3}}^{(3)} & = & 1 ,
\nonumber \\ 
{\cal P}_{\rho_{1} \rho_{3} \rho_{2}}^{(3)} & = & 
3\sqrt[3]{\lambda_1 \lambda_2 \lambda_3}  , 
\end{eqnarray}
where the latter requires full rank to be nonvanishing and we 
have used $(-1)^{N-1} \det U^{\parallel} = +1$ for $N=3$. 

Let us now turn to the issue how to measure 
$\gamma_{\rho,\rho^{\perp}}^{(2)}$. In general, the procedure 
to achieve this is based upon lifting $\rho$ and $\rho^{\perp}$ 
to the pure states $\ket{\Psi_{sa}}$ and $\ket{\Psi^\perp_{sa}}$, 
respectively, by attaching an ancilla system $a$ in such a way 
that $\rho = \Tr_a \ket{\Psi_{sa}} \bra{\Psi_{sa}}$ and 
$\rho^{\perp} = \Tr_a \ket{\Psi_{sa}^{\perp}} \bra{\Psi_{sa}^{\perp}}$. 
This purification can be performed experimentally in the qubit 
case by using a Franson type interferometer set up  
and polarization-entangled photon pairs (system and ancilla photon) 
\cite{franson89}, see Fig. 2. A source of this type that produces 
photons in the horizontal-vertical ($h-v$) basis has been 
demonstrated in \cite{kwiat99}. Such photon pairs are sent 
to the interferometer in the polarization-entangled state 
$\ket{\Psi_{sa}} = \sqrt{\frac{1}{2}(1+r)} \ket{h} 
\otimes \ket{h} + \sqrt{\frac{1}{2}(1-r)} \ket{v}\otimes\ket{v}$ 
and in the short arms the polarization is flipped so as to 
obtain $\ket{\Psi_{sa}^{\perp}} = \sqrt{\frac{1}{2}(1-r)} \ket{h} 
\otimes \ket{h} + \sqrt{\frac{1}{2}(1+r)} \ket{v}\otimes\ket{v}$, 
where $r$ is the degree of polarization in single-photon 
measurement. By an appropriate choice of unitarity 
$V$, the off-diagonal phase 
$\gamma_{\rho \rho^{\perp}}^{(2)}$ can be measured as a 
shift of the interference pattern obtained by varying the 
$U(1)$ phase $\chi$ in this two-photon scenario. For simplicity, 
one may consider parallel transporting unitarities that rotates 
linear polarization into linear polarization. With $\beta$ the 
polarization angle of the photons with respect to the horizontal 
axis, this amounts to $U^{\parallel} = \exp \big[ -\beta 
(\ket{h}\bra{v} - \ket{v}\bra{h})\big]$, and the desired 
output intensity detected in coincidence is 
\begin{eqnarray}
\label{eq:int4}
{\cal I} & \propto & \big| e^{i\chi} \ket{\Psi_{sa}^{\perp}} + 
U^{\parallel} \otimes V \ket{\Psi_{sa}} \big|^{2} 
\nonumber \\ 
 & \propto & 1 + \big| \Tr \big( U^{\parallel} 
\sqrt{\rho} U^{\parallel} \sqrt{\rho^\perp} \big) \big| 
\cos \big( \chi - \gamma_{\rho,\rho^{\perp}}^{(2)} \big)
\end{eqnarray} 
if we choose $V = \exp \big[ \beta (\ket{h}\bra{v} - 
\ket{v}\bra{h})\big]$. Explicit calculation predicts  
$\Tr[\sqrt{\rho} U^{\parallel} (\beta) \sqrt{\rho^\perp} 
U^{\parallel} (\beta) ] = \sqrt{1-r^2} \cos^{2} \beta - 
\sin^{2} \beta$, which can be positive and negative for 
$r\neq 1$ depending upon $\beta$. Thus, such an experiment 
would test that the off-diagonal geometric phase is either 
$0$ or $\pi$ for mixed qubit states. 

\begin{figure}[ht!]
\begin{center}
\includegraphics[width=8 cm]{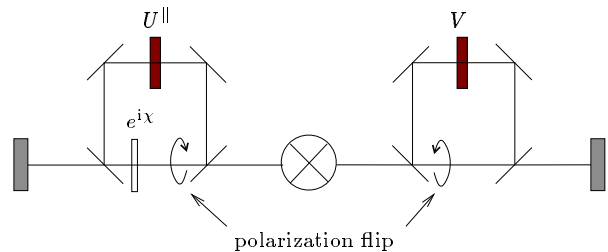}
\end{center}
\caption{Franson set up for polarization-entangled photon pairs. 
Polarization flips are applied to the shorter arms. In the longer 
arms, the photons are exposed to the unitarities $U^{\parallel}$ 
and $V$, respectively. The shift in the coincidence interference 
oscillations obtained by varying the $U(1)$ phase $\chi$ is 
determined by the pair of unitarities in the longer arms.}
\end{figure}

In conclusion, we have introduced the concept of off-diagonal 
geometric phase for mixed states. In the qubit case we have 
demonstrated that the nodal points of the diagonal and 
off-diagonal mixed state geometric phase never coincide. 
Extension to cyclic products of density 
matrices in arbitrary Hilbert space dimensions is shown 
to further enrich the mixed state phase. We have also proposed 
a polarization-entangled two-photon experiment that could test 
the off-diagonal mixed state geometric phase, and in particular 
check the sign change property across its nodal surfaces. 

E.S. acknowledges financial support from the Swedish Research 
Council.

\end{document}